\documentclass[%
 reprint,
 amsmath,amssymb,
 aps,
]{revtex4-1}

\usepackage{graphicx}
\usepackage{dcolumn}
\usepackage{bm}
\usepackage{xcolor}
\usepackage{soul}

\begin{document}


\title{Coherent control of light for non-line-of-sight imaging}

\author{Ilya Starshynov$^*$}
\author{Omair Ghafur}%
\author{James Fitches}%
\author{Daniele Faccio}%
\thanks{ilya.starshynov@glasgow.ac.uk; daniele.faccio@glasgow.ac.uk}
\affiliation{School of Physics \& Astronomy, University of Glasgow, Glasgow G12 8QQ, UK 
}%

\date{\today}

\begin{abstract}
Non-line-of-sight (NLOS) imaging relies on collecting light that is rendered incoherent from the multiple scattering events and is then post-processed to provide an estimate of the hidden scene. Here we employ coherent phase control of the outgoing laser beam phase front so as to refocus the beam behind the obscuring obstacle and then use the speckle memory effect to scan the focused spot across the scene. The back-reflected light intensity provides a direct measurement of the scene with a signal-to-noise ratio that is greatly improved when measured using a temporally gated detector. A spatial resolution of less than 1 mm is demonstrated, opening the way to high-resolution NLOS imaging.   
\end{abstract}

\maketitle


{\bf{Introduction.}} Imaging objects not accessible to direct observation a.k.a non-line-of-sight (NLOS) imaging is a challenging problem with a variety of applications in bio-medical imaging, autonomous robotics, defence and security \cite{review}. There exist a large number of proposed reconstruction techniques~\cite{freund1990looking,kirmani2009looking,Naik:14,velten2012recovering,o2018confocal,bertolotti2012non,katz2012looking,katz2014non,gariepy2016detection}, which can be split into two major categories: those exploiting the time of flight (TOF) of the photons from the source to the hidden object~\cite{kirmani2009looking,Naik:14,gariepy2016detection,o2018confocal} and those based on correlations of the intensity fluctuations of coherent scattered light encountering the object ~\cite{freund1990looking,bertolotti2012non,katz2012looking,katz2014non}. Although both of these categories require a diffusive scattering surface, the conditions of their applicability are different. 
The resolution of TOF methods is limited by the detector performance. The state-of-the-art single photon avalanche photodiodes (SPAD) offer 30 ps temporal resolution \cite{sanzaro2017single}, which corresponds to a path difference of around 1 cm, and depending on the experiment geometry allows to resolve 1-10 cm features~\cite{velten2012recovering,o2018confocal}.

In the case of coherent light, diffusive scattering from a rough surface leads to random interference patterns called speckle whose intensity distribution can be controlled or even refocused by shaping the wavefront of the incident field~\cite{vellekoop2007focusing}. Moreover, the so-called speckle memory effect~\cite{FreundMem,Feng} implies that changing the incidence angle of the incoming wave causes the same tilt of the output speckle without changing its shape, assuming the tilt is not too large. 
\begin{figure}[t!]\label{setup}
\centering
\includegraphics[width=6.5cm]{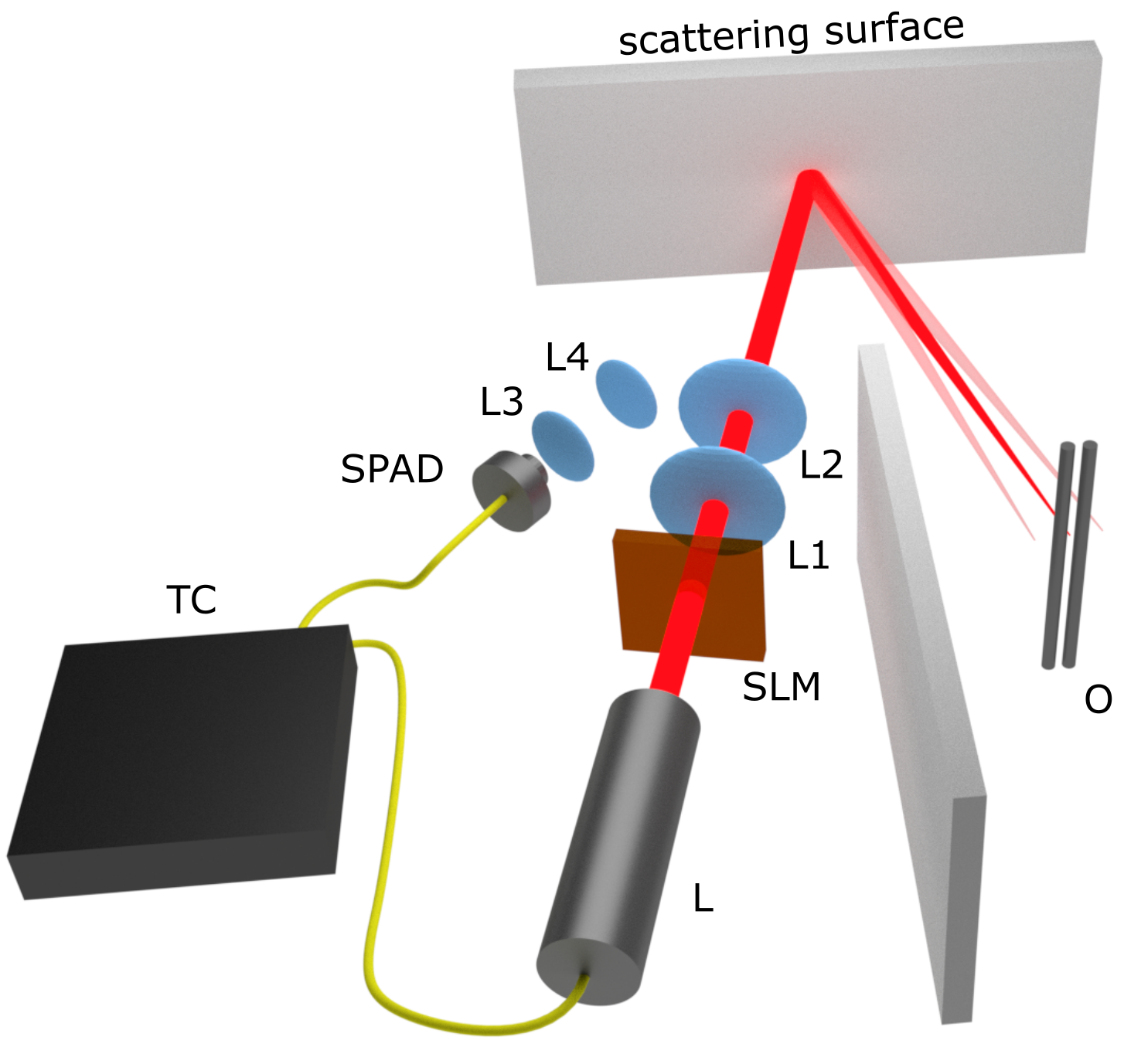}
\caption{Experimental setup. A pulsed laser is shaped with a SLM, which is imaged onto a scattering surface. First, the SLM pattern is optimized to achieve a re-focused spot in the scattered field and an additional phase gradient is added to scan the spot. An image of the object placed in the scanning region is reconstructed by measuring the intensity the light reflected from the object after reflection from the wall again.}
\end{figure}
 Imaging methods based on this effect allow diffraction-limited reconstruction within the field of view limited by the range of the memory effect (up to 1$^\circ$). However, the information about the object shape is encoded into small intensity variations of the raw data image against a background that is proportional to the total object luminosity. This does not allow to illuminate the object with the light diffusively scattered from the same rough surface due to the strong background signal, also originating from the first reflection from the surface. 
Therefore, any of the speckle memory effect-based techniques known to the authors require a self-luminous object. In addition, the illuminated area on the rough surface increases when the object is moved away from it, which leads to smaller features in the raw data images and places constraints on the distance of the imaging sensor to the rough surface (10 cm in~\cite{katz2014non}).

Here we combine memory effect and TOF based techniques to obtain high resolution images of an object placed behind an occluding obstacle by coherent control of the light scattered from a rough reflective surface. We build a theoretical model that describes the main features of this configuration relevant to the choice of the experiment parameters such as the angle of incidence and optimal distance from the surface based on the physical parameters of the scattering medium. We use wavefront shaping to experimentally focus scattered light into a spot, which we are then able to scan within a particular region using the speckle memory effect.This allows to achieve sub-mm spatial resolution, thus significantly improving time-of-flight techniques.
\newline

{\bf{Refocusing diffuse light for NLOS imaging.}}
The setup is illustrated in Fig.~1. As a source of light we use a Ti:sapphire pulsed laser (coherent Chameleon Ultra) at the wavelength $\lambda$ = 810 nm. The laser beam is expanded to fill the surface of a reflective spatial light modulator (SLM, Meadowlark  512$\times$512 pixels,  7.68$\times$7.68 mm active area), which is imaged onto a scattering surface (brushed aluminium plate) with a demagnification factor of 0.8. In order to create a spot in the reflected speckle we put a camera at the supposed position of the object and perform an optimization by measuring the reflection matrix of the surface. The light back scattered from the object when this is put in place of the camera is collected from a 3.5 cm spot on the scattering wall by means of a telescope focusing it to a SPAD. TOF histograms are collected at 15 Hz repetition rate using a time-to-digital (TDC) converter (ID900  IDQuantique) synchronized with the laser. 
\begin{figure}[!t]\label{fig:enh}
	\centering
	\includegraphics[width=0.45\textwidth]{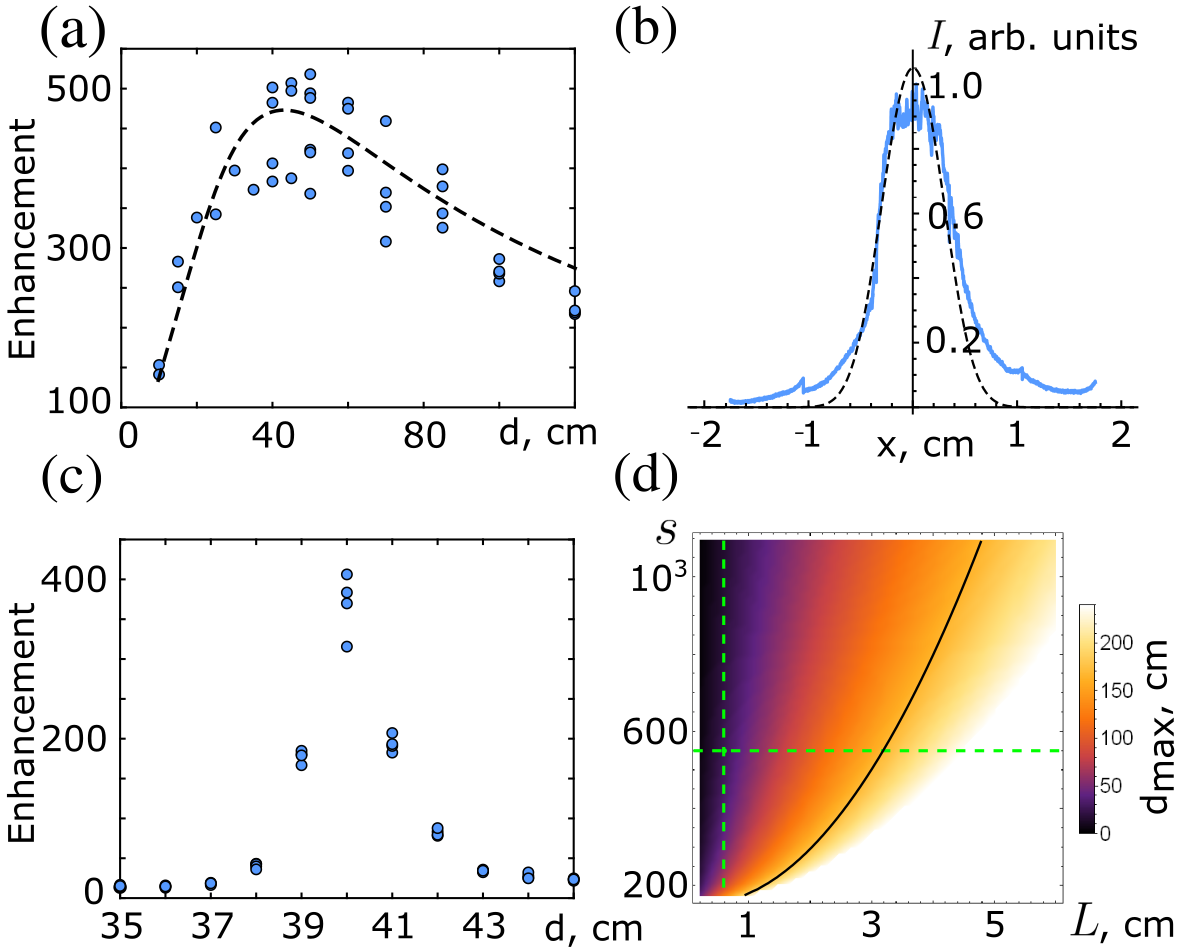}
	\caption{a) Dependence of the spot enhancement factor (ratio of the focused spot intensity to the average background intensity) on the distance from the scattering surface, $d$, performing re-optimization of the SLM pattern at each distance. For each $d$ multiple measurements are performed changing the illumination position on the surface. Dashed curve is a fit of $N(d)$ from Eq.~(\ref{eq:N_d}) with the only fitting parameter $N_0\approx 10^5$. b) Average intensity of the reflected speckle at 40 cm from the scattering wall, from which we determine $s\approx550$ (dashed fitted curve). c) Dependence of the spot enhancement on the distance from the scattering surface while keeping a fixed SLM pattern. d) Dependence of the $d_\text{max}$ on the size of the illumination beam at the surface,~ $L$ and $s$ from Eq.~(\ref{eq:N_d}) }
\end{figure}
In most of the experiments reported in literature related to scattered light focusing, the incoming light is incident normally onto the scattering sample. However, since the speckle patterns become highly anisotropic at large scattering angles~\cite{Goodman}, in our experiment we tilted the incident wavefront by $\sim$6-10$^\circ$ with respect to the scattering surface.

{\bf{Theoretical model.}}
{We wish to refocus the laser power into a spot at macroscopic distances, e.g. 10-100 cm from the scattering surface. The intensity enhancement of the refocused spot will be proportional to the number of modes at a given distance. We expect little enhancement very close to the surface as speckle forms due to interference between various plane wave components that occur only upon propagation away from the surface. On the other hand, in the limit of infinite distance, the speckle produced by the surface turns into a plane wave according to the Van Cittert-Zernike theorem~\cite{Goodman}, i.e. again, only one mode participates to the light intensity at any given point. At these two opposite regimes, no speckle refocusing (intensity enhancement) can be expected, implying an optimal enhancement somewhere in between. 

We first note that the angular spread of the reflected beam is described by~\cite{bass1979wave}
\begin{align}
	& \Omega(\psi,\theta,\phi) =\frac{(1-\cos (\theta) \cos (\phi) \cos (\psi)+\sin (\theta) \sin (\psi))^2}{(\sin (\theta)+\sin (\psi))^2}\times \nonumber \\
	 &\exp \left(-\frac{\cos ^2(\theta)+\cos ^2(\psi)-2 \cos (\theta) \cos (\phi) \cos (\psi)}{4 s^2 (\sin (\theta)+\sin (\psi))^2}\right),
\end{align}
where $\Omega(\psi,\theta,\phi)$ is the relative amount of emission at the direction given by the inclination angle $\theta$ and azimuth angle $\phi$ assuming the rough surface is illuminated by a plane wave incident at an angle $\psi$, and $s$ is the ratio between the average height and width of the surface roughness features. Assuming we illuminate a square area of a rough surface with a side $L$ with a normally incident plane wave, the number of reflected modes directed at a particular point in space is
\begin{align}\label{eq:N_th}
    N_{\theta}(x,y,z) = &\frac{N_0}{L^2}\int \int_{-L/2}^{-L/2} \Omega\left(0,\theta(x+v,y+u,z), \right. \nonumber\\ &\left.\phi(x+v,y+u) \right) du dv.
\end{align}

In the case of a rectangular illuminated area, the average width of the speckle grain changes with the distance from the surface as $\lambda z/L$, which leads to effective reduction of the number of modes with $z$ at this rate. The resulting dependence of the number of modes on the distance from the surface $d$ at a point $(x_0,y_0)$ is therefore
\begin{equation}\label{eq:N_d}
	N(d) = N_0 N_\theta(x_0,y_0,d)/L\lambda d.
\end{equation}

Equation~(\ref{eq:N_d}) has a maximum at a particular distance $d_\text{max}$ indicating that indeed, for a given scattering surface (described by $s$) and beam size, there is an optimal refocusing distance.  We experimentally verified this by refocusing the laser spot (method described below) and measuring the dependence of the focused spot enhancement on  $d$, see Fig. 2(a). 
As can be seen, the maximum enhancement (that is specific for the surface we use) is reached at a distance $d_\text{max}\sim$40 cm. In Fig. 2(a) we also show the fit of the theoretical dependence $N(d)$, Eq.~(\ref{eq:N_d}) to our data, where the only fitting parameter is the normalisation constant $N_0$, while $s$ is determined by fitting { $N_{\theta}$, Eq.~(\ref{eq:N_th}), which is proportional to the average intensity distribution of the scattered light, to an experimentally measured distribution, see Fig. 2(b)}}. We performed all further experiments placing the object at this optimal distance of 40 cm. Figure~2(c) shows the experimentally measured depth of focus of the refocused spot, i.e. we fix the SLM phase pattern and then move the camera from 35 to 45 cm measuring the focused spot enhancement. 

Figure 2(d) shows the full dependence of the optimal focusing distance $d_\text{max}$ as a function of the surface parameter $s$ and beam size. For a given diffusive surface, one can tune the optimal refocusing distance by changing the beam size, however, small $s$ or large beam sizes also lead to a reduction in the actual peak intensity enhancement: the solid black curve indicates the values for which the enhancement, wrt to the surrounding speckle, reduces to a factor of 100$\times$, compared to 500$\times$ obtained for the surface we use. 
	\begin{figure}\label{fig:spots}
	\centering
	\includegraphics[width=0.45\textwidth]{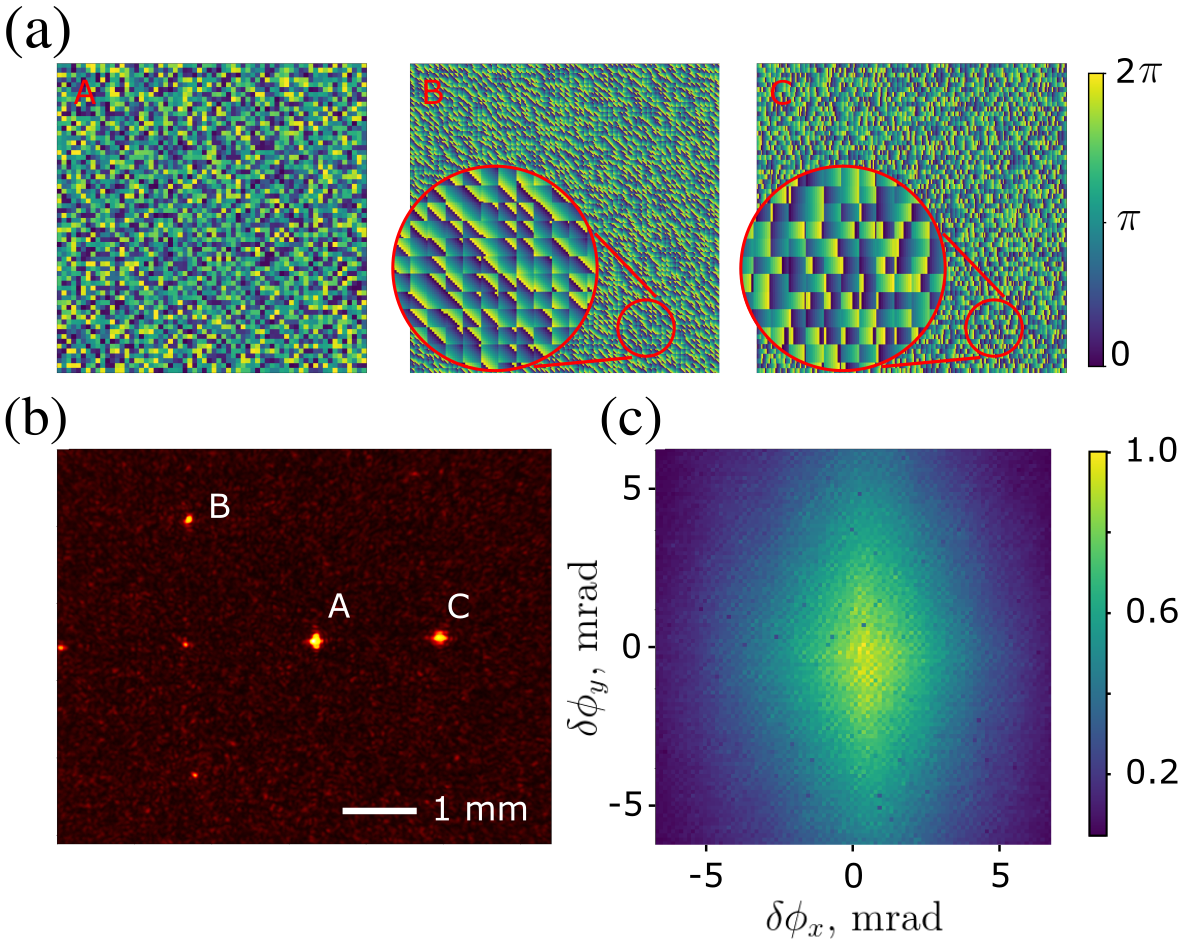}
	\caption{Manipulating the scattered light: a) phase patterns on the SLM that correspond to the focused spots A, B and C in the reflected speckle pattern shown in b). {Additional unlabeled bright spots are artifacts from the diffraction on the macropixels} c) Attenuation of the focused spot relative to zero displacement. }
	\end{figure}

{\bf{Optimization method.}}	
There exist a variety of optimization methods~\cite{vellekoop2015feedback}, starting from the most simple continuous-sequential and partitioning algorithms~\cite{vellekoop2007focusing,vellekoop2008phase}, to more advanced genetic~\cite{Conkey:12} or simulated annealing~\cite{fang2018focusing} algorithms. 
Alternatively, some methods exploit the physical properties of the scattering process, in particular its linearity. 
The output modes are related to the input modes by means of a scattering matrix~\cite{Beenak}. 
Knowledge of this matrix allows to design a configuration of the input modes required to achieve (almost) any desired output intensity distribution. Therefore the optimization task reduces to measuring the scattering matrix of the disordered sample~\cite{Popoff_2011}. In practice, the major limiting factor defining the performance of any of these algorithms is the measurement noise~\cite{vellekoop2015feedback}. 
 In the case of bulk scattering media, the main source of speckle instability is sample alteration. As the speckle pattern is highly sensitive to the position of every single microscopic scatterer in the disordered material, it changes  with temperature, humidity and even ambient pressure for such samples~\cite{albertazzi2018speckle}. In our case the scattering surface configuration remains stable over days, and mechanical vibrations become the major source of instability. These deteriorate the performance of the algorithms which optimize the input pattern based on a single output spot intensity.
The scattering matrix measurement does not rely on a single output spot and is therefore less affected by the short-scale mechanical sample instability. For this reason we use the scattering (reflection) matrix measurement to focus the scattered light in our experiment.

We measured the reflection matrix of the scattering surface using the internal reference method~\cite{popoff2010measuring}. This method is a variation of a standard interferometric technique in which a set of phase patterns is selected and then each of them is shifted in 4 steps of $\pi/2$ with respect to a static reference while recording the resulting intensity distribution. In our case, the static reference originates simply from light reflected from the areas on the SLM not covered by the liquid-crystal pixels. 

We use 64$\times$64 Hadamard patterns as the basis and capture the intensity on a 100$\times$100 pixel region of the camera for each of the 64$\times$64$\times$4 measurements, from which the resulting reflection matrix is reconstructed. The input phase distribution leading to the focused spot in the output is then obtained as a product of the inverse of the measured matrix with the desired intensity image. We approximated the inverse of the reflection matrix by its Hermitian conjugate, effectively performing phase conjugation, since this method is more resilient to the measurement noise~\cite{popoff2010measuring}.
	
{\bf{Scanning the focused spot.}}
 The knowledge of the scattering matrix allows to get a focused spot at any of the 100$\times$100 output positions. However, scanning the spot using the scattering matrix would make our method conceptually invasive, wile the memory effect allows to move the focus outside the region in which the scattering matrix was measured. Therefore, after we obtain a focused spot in the output speckle we displace it by adding a linear phase gradient to the wavefront incident on the SLM, see Fig.~3. In our experiment we were able to shift the speckle spot by several millimeters (see Fig.~3(b,c)). However the main limiting factor was not the memory effect range, but rather the dimensions of the SLM. One full phase gradient from 0 to 2$\pi$ across one of the dimensions of the SLM (7.68mm) corresponds to a tilt of 0.11 mrad, which corresponds to a displacement of $\sim$42 $\mu m$ at 40 cm from the scattering surface. In order to achieve large displacements, the phase has to wrap many times across the SLM, as can be seen from Fig.~3(a), until the limited spatial resolution can no longer correctly represent the linear gradient. We checked the full memory effect range by physically tilting the SLM and found that it is possible to achieve a displacement of $\pm$1.5$^\circ$ loosing only $50\%$ of the spot intensity, corresponding to a shift of {{$\pm$~1~cm}} at 40 cm distance from the wall. This indicates that our method can be used to focus light on one side of the barrier and then scan it at the other side by a properly applied phase gradient. 
\begin{figure}[t]\label{fig:res}
	\centering
	\includegraphics[width=0.45\textwidth]{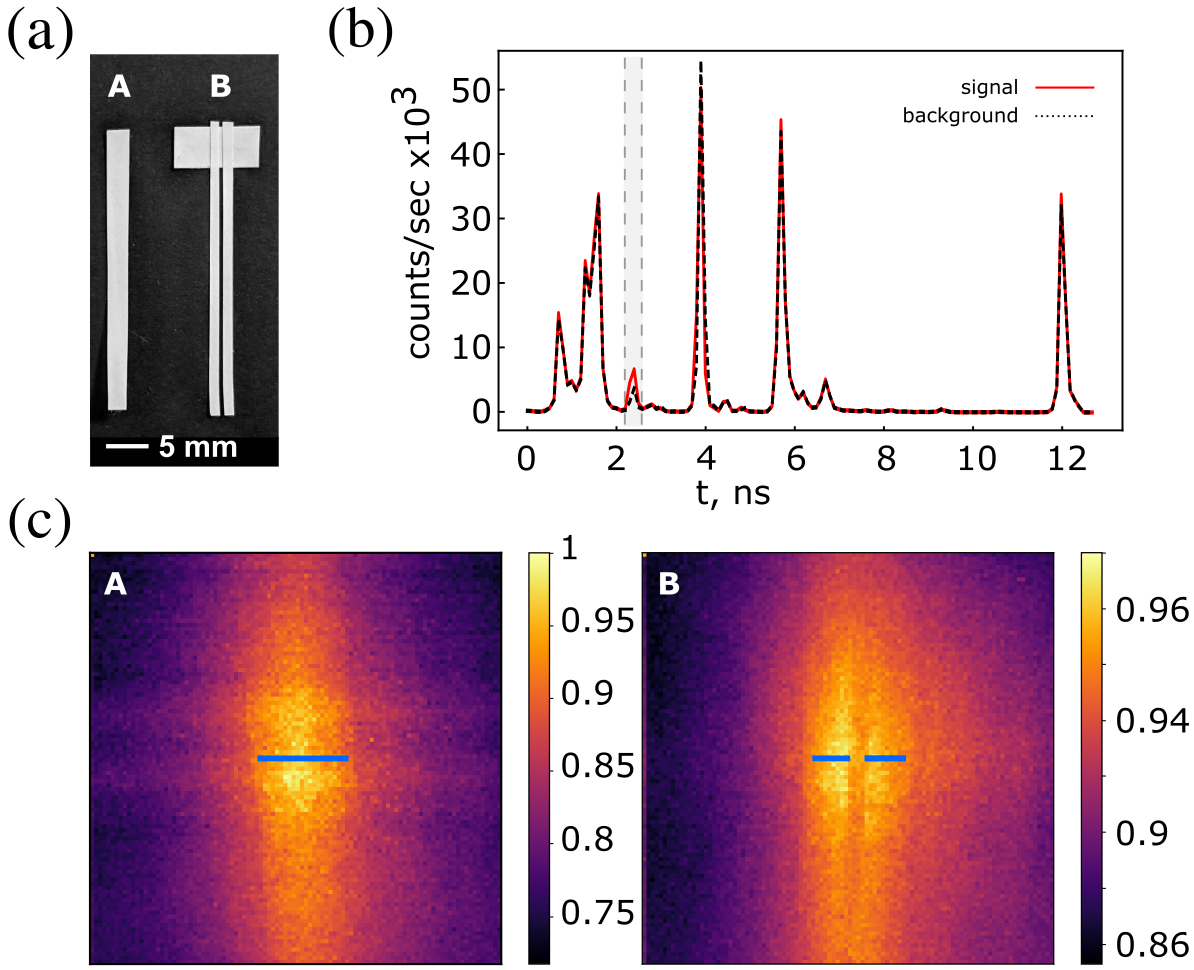}
	\caption{a) Two objects that we imaged with our method. b) Time-of-flight histograms measured with the SPAD: red solid line, the focused spot impinges on the object; black dashed line, without any re-focusing (flat pattern on the SLM). The shaded area shows the temporal (time of flight) delay corresponding to the object position. c) Measured images of the objects A and B in a). Blue lines show the actual size of the objects (2.1 mm, object A, 2.8 mm object B)}.
\end{figure}

{\bf{Results.}}
After optimizing the focused spot we replaced the camera with an object (see Fig.~4(a)) and performed a 100$\times$100 points scan with a 2.34 mrad (93 $\mu$m at 40 cm) step by adding corresponding phase gradients to the SLM pattern as discussed above. 
In each of the focus positions we measured the time-of-flight (TOF) histogram.
Two example TOF histograms of the light collected at the scattering surface are shown in Fig.~4(b): the red solid line is the histogram recorded while the focused spot was hitting the object, the black dashed line shows the background obtained by putting a flat phase pattern on the SLM and thus eliminating the focus. The difference between these two graphs contains information about the object reflectance and more importantly, allows to identify the TOF peak (at 2.2 ns) that corresponds to reflection from the object, as it is the one showing the highest variation between the two measurements. We then integrate the temporal histogram data only in a small (0.5 ns) window around this peak. This procedure provides a very effective temporal filter and rejection of background counts, increasing the sensitivity of our method. 

The scans of the object A and B obtained by this approach are presented in Fig.~4(c). In these experiments we used laser power of only around 100 mW, which leads to a return signal from the object of around 6500 counts/sec. Such a variation corresponds to only  1.5 \% of the total detected light intensity, which would make it difficult to reconstruct the object using any of the speckle memory effect based methods~\cite{bertolotti2012non,katz2012looking}. On the other hand conventional TOF imaging techniques would also fail in this scenario as these rely on detecting the difference in the time of arrival of photons emitted from different points on the object. In our geometry (object at 40 cm from the wall) the maximal difference in the distance to the detector for two points separated by 100 $\mu$m, is $\sim$ 70 $\mu$m, which corresponds to a time delay of 0.23 ps, which is far beyond the resolution of any existing time-resolving detector. The resulting images, however, have less contrast towards the edges. Effectively the reconstructed image is a product of the object's actual shape and the dependence of the focused spot intensity on the displacement angle, Fig.~3(c). 

{\bf{Conclusion.}}
We have demonstrated a novel imaging technique based on the combination of the TOF reconstruction and speckle memory effect based imaging. Our technique implies focusing light reflected from a rough surface into a single spot by shaping the incident wavefront, which then can be displaced using the memory effect. Imaging is achieved by scanning the spot within a particular area and recording TOF histograms of the return light. The resolution of our method is defined by the size of the focused spot, which can be minimized potentially to the diffraction limit, with the only limitation being the decrease of the total energy within the focus. The time resolved measurement allows to detect small changes in the total output signal due to the presence of the object, which makes our method applicable in the situations where other memory effect based imaging techniques do not work. Although in the current work we use a camera in place of the object to measure the scattering matrix and focus the reflected light prior to the object reconstruction, the memory effect gives us knowledge about the part of the scattering matrix we never measured. We show that its range can be enough for fully non-invasive reconstruction, that can be achieved by focusing at one side of the obstacle and successively displacing the spot behind it. In the current experiment we couldn't achieve that because of the SLM not being able to apply a steep linear gradient accurately. This limitation could be eliminated by using, for example, a galvanometric mirror in combination with the SLM.

{\bf{Acknowledgements.}} 
This work was supported by the Engineering and Physical Sciences Research Council of the UK (EPSRC) Grant numbers EP/M01326X/1, EP/S026444/1, the UK MOD University Defence Research Collaboration (UDRC) in Signal Processing and the UK MOD Future Sensing and Situational Awareness Programme.


%

\end{document}